# Light Fidelity in Academic Libraries: A Radiant Journey to Light the Future

JAGAN B L
Research Scholar
Dept. of Library & Information Science,
University of Kerala, Thiruvananthapuram
https://orcid.org/0009-0007-2295-797X
jaganbl2024@keralauniversity.ac.in

Dr. SUDHI S. VIJAYAN
Assistant Professor & Head,
Dept. of Library and Information Science,
University of Kerala, Thiruvananthapuram,
ORCID: 0000-0002-2228-9031,
sudhivijayan@keralauniversity.ac.in

Dr. V. R. RENJITH
Librarian,
Dept. of History,
University of Kerala, Karyavattom Campus,
Thiruvananthapuram.
ORCID: 0000-0002-2389-7338,
renjithlib@keralauniversity.ac.in

Dr. ARUNKUMAR V.R.
Assistant Librarian Gr. II,
Kerala University Library, University of Kerala, Thiruvananthapuram.
ORCID: 0000-0001-7344-6263,
arunkumarvr@keralauniversity.ac.in

## Abstract

Academic contexts have changed due to the fast expansion of communication technology, affecting information access, sharing, and dissemination. Light Fidelity (LiFi) is a new technology that may transform academic library connections in this changing environment. This article examines how LiFi can accelerate data transmission, secure data, and provide a more dynamic and collaborative learning environment to spur creativity. Academic libraries may accelerate information interchange using visible light communication with LiFi technology. LiFi is the solution for the increasing need of digital access it adress the situation by enabling faster research activities and enhancing collaboration among students, faculty, and researchers. Present paper discusses the importance and application of LiFi in libraries.

## 1. Introduction

Communication technologies promote global connection through the transmission of information between people and equipment. These technologies range from established techniques like radio waves and wired networks to state-of-the-art inventions that always influence how we communicate share information, and work together in our connected world. In the present AI technological world, the current standard Wireless Fidelity (Wi-Fi) Communication-based technology will not be able to meet the desired data requirement. Adding Light Fidelity (LiFi) to traditional Wi-Fi-based systems can be an extremely successful addition (Verma et al., 2023). In the absence of Wi-Fi, which transmits data via radio waves, LiFi utilizes Visible Light Communication (VLC), indicating a paradigm shift in wireless connectivity. LiFi effectively converts standard LED lamps into data transmission sources using the visible light spectrum. LiFi operates on the modulation of light, rendering it a more adaptable and secure alternative to WiFi, which utilises radio frequency signals. LiFi is capable of transmitting data at speeds that are far faster than those of Wi-Fi. The limited nature of light leads to improved security, which reduces the possibility of data interception. It leverages pre-existing lighting infrastructure to promote energy savings and environmental behaviours. In the context of academic libraries switching to the digital age, it is critical to develop an in-depth knowledge of LiFi to facilitate secure and efficient wireless communication in the future.

## 2. Components of LiFi

LiFi, or Light Fidelity, transmits data using visible, infrared, and ultraviolet light. LiFi transmits data via light waves. The main LiFi components are:

1) **LED Light Source:** LiFi technology relies on LED light sources. LEDs are used for light emission, and the luminosity of these light signals may be adjusted to transmit data.
2) **LED driver:** LED drivers adjust light intensity. The LED brightness may be adjusted in real-time for data modulation.
3) **Photodetector:** Photodiodes or arrays detect modulated light signals. Light signals are converted to electrical impulses by the photodetector.
4) **Circuits for modulation and demodulation:** Modulation encodes data into LED light pulses. Demodulation reverses light signals to obtain encoded data. Both transmitter and receiver communication depend on these circuits.
5) **Signal Processing Unit:** The signal processing unit processes photodetector electrical signals. It may entail error correction, signal amplification, and conditioning.
6) **LiFi Transceiver:** The LiFi transceiver transmits and receives in a single device. The LED light source, photodetector, and modulation/demodulation circuits enable bidirectional communication.
7) **Communication Protocol:** Communication protocols manage data transmission, reception, and management in LiFi. IEEE 802.15.7, a popular protocol, defines the physical layer and medium access control for visible light communication.
8) **LiFi Access Points:** Like WiFi routers, LiFi access points link numerous LiFi-enabled devices locally. They control device data transfer and network management.
9) **LiFi-Enabled Devices:** Devices like smartphones, tablets, laptops, and other gadgets with LiFi transceivers can communicate using LiFi technology.

## 3. Benefits of LiFi in Academic Libraries

LiFi technology is a light of innovation in academic libraries, providing several benefits beyond standard connection. Academic libraries always seek quicker and more efficient data transport. LiFi promises faster speeds and better connections. This discussion will explore the potential of LiFi's lightning-fast speeds to transform how patrons utilize online databases and digital resources in academic libraries. Human safety is the main benefit of VLC wireless networking. Humans constantly absorb electromagnetic energy since wireless routers, Bluetooth gadgets, and similar wireless systems continuously release it (Shaaban & Faruque, 2020).

### 3.1 High-Speed Connectivity

VLC systems have traditionally worked as simple point-to-point wireless connections between an LED light and a receiver with a photodetector, and how fast data moves through these systems comes down to the modulation method and lighting tech being used (Hass, 2018). LiFi really stands out when it comes to its ability to transmit data at lightning-fast speeds. This will result in quick access digital resources, smooth downloading of research papers and retrieval of information instantly. For faculty, students, and researchers, this speed is a game-changer, allowing them to sift through large amounts of information more efficiently than ever before. LiFi uses visible light to trasmit data which makes it much faster than the traditional WiFi. This speed greatly benefits libraries with majority of digital resources, research databases, and online journals. With LiFi, accessing digital libraries, eBooks, multimedia content, research papers, and teaching videos becomes effortless, enhancing the learning experience and allowing for quicker progress in your studies. Say goodbye to frustrating delays and hello to a more efficient way to explore knowledge! Modern academic research depends heavily on people working together from different places, and LiFi's fast connection means researchers can share big datasets, hold virtual meetings that don't lag, and collaborate on projects as things happen in real-time, which makes research better and sparks more creative thinking. Academic libraries deal with heavy data use all the time—lots of people accessing resources at once, big files moving around, videos streaming—and regular wireless networks get clogged up under this kind of pressure, but LiFi's speed cuts through these logjams so people can get to digital resources without hitting slowdowns or annoying waits while they're studying or doing research.Claude is AI and can make mistakes. Please double-check responses.

**3.2 Enhance Security**

Security matters enormously in academic settings where sensitive research data gets shared regularly, and LiFi's restricted signal range—which only works in lit areas adds a natural security layer by reducing the chances of sednarios like data breaches and unauthorized access. In an era where information integrity is critical, LiFi acts as a guardian of academic data by combining speed with robust security that protects against various threats. LiFi works within the limits of visible light, that is transmission can only reach the areas which are illuminated unlike WiFi that pass through walls, and this geographical limitation protects important research and user data in academic libraries from unauthorized access. Electrical equipment that emits radio frequency can interfere with WiFi networks. While LiFi avoids this problem by using light exclusively, it makes data transmission more reliable and eliminates disruptions that could compromise academic information. Eaves dropping and interception can be compltetly avoided with the physical barrier like wall which naturally limit the visible light technology that LiFi uses. Cyber attacks threats are minimal because of the LiFi inherent limitation the hackers need to remotely access the trasmissions remotely. Along with these physical security advantages, LiFi networks can implement advanced encryption that makes intercepted data useless without proper decryption keys, adding another layer of protection for sensitive research findings and personal information stored in academic databases.

**3.3 Energizing Sustainability in Academic Libraries**

Academic libraries should embrace LiFi and sustainability. Using existing lighting infrastructure for data transmission eliminates the demand for networking equipment and saves energy. Typically, light signals are modified using light-emitting diodes (LEDs) to transfer data. LiFi systems save energy and are greener than other connectivity solutions. Lower power consumption than Wi-Fi is usual with LiFi. This is because data transmission LEDs may be muted or switched off when not in use, saving energy. Visual light does not interact with other wireless technologies; hence, LiFi may operate in this spectrum. It may improve electromagnetic spectrum usage, avoiding congestion and reducing interference power. LiFi's environmental benefits and how they affect academic library sustainability are discussed below:

- **Utilizing Existing Infrastructure for Dual Functions:** LiFi uses lighting infrastructure for data transmission. Academic libraries may use LED lights as data transmitters. It optimizes resource utilization and eliminates the demand for networking equipment, boosting energy efficiency.
- **Reduced Energy Consumption Compared to Wi-Fi:** Energy Savings LiFi uses less energy than Wi-Fi. Wi-Fi utilises radio frequency, whereas LiFi uses visible light. Energy-efficient LED lighting makes LiFi installations greener. Academic library operations are more sustainable and environmentally friendly with energy conservation.
- **Smart Lighting Control for Enhanced Efficiency:** LiFi provides smart lighting system management. Libraries may adapt lighting depending on occupancy, time of day, or natural light using smart lighting management. It improves energy efficiency and makes the library more user-friendly.
- **Lower Electromagnetic Interference (EMI):** LiFi's use of visible light reduces EMI, a frequent issue with radio frequency systems like Wi-Fi. Reduced EMI improves electronic equipment dependability and supports sustainable practices by minimizing library electronic system disturbances.

- **Green Initiatives and Corporate Social Responsibility (CSR):** Adopting LiFi technology shows a commitment to green initiatives and corporate social responsibility (CSR). Many educational institutions are prioritizing sustainability. The library's use of LiFi shows its commitment to environmental management and carbon reduction.
- **Low Environmental Impact:** LiFi's visible light connectivity removes the requirement for radio frequency infrastructure. It decreases the environmental effect of typical Wi-Fi equipment manufacture, deployment, and disposal. Academic libraries that use LiFi promote greener technology.
- **Aligning with Campus Sustainability Goals:** Academic libraries often operate within sustainability-focused campus contexts. LiFi's energy-efficient networking supports these campus-wide activities. Academic libraries may support campus sustainability by working with sustainability teams.

### 3.4 Enhances Smart Libraries: IoT Integration for Connectivity and Intelligence

The integration of LiFi and IoT devices represents a paradigm shift in academic libraries, transforming them into intelligent, connected ecosystems that extend far beyond basic connectivity. The combination of LiFi with smart lighting systems will help the LED lights to perform dual purposes that is along with illumination the lights can be used to transmit data, this will make the connectivity in the library strong. The collaboration enables precise location-based services through IoT sensors, allowing patrons to access real-time positioning information, resource discovery tools, and location-triggered alerts for a more immersive user experience. Furthermore, intelligent sensor networks facilitate effective resource management by monitoring book inventories, study room availability, and equipment usage, empowering administrators to make data-driven decisions and optimize resource allocation. The integration also enhances responsive environments through smart displays and interactive kiosks that dynamically adapt content based on user preferences, ongoing events, and real-time data, while enabling seamless real-time collaboration tools that allow students and researchers to effortlessly connect devices for sharing documents and virtual workspaces. Ultimately, this LiFi-IoT convergence promotes energy-efficient and sustainable practices, as libraries leverage LED infrastructure for both lighting and data transmission, reducing overall energy consumption and reinforcing the institution's commitment to environmentally responsible operations.

## 4. Applications and Implications of LiFi in Academic Libraries

In the dynamic landscape of academic libraries, integrating Light Fidelity (LiFi) technology represents a transformative leap in connectivity. This comprehensive overview delves into LiFi's myriad applications and profound implications for academic institutions, redefining how knowledge is accessed, shared, and safeguarded.

- **Faster Data Transfer for Digital Resources:** LiFi is known for its exceptional data transfer rates. Digital resources, scientific publications, and internet databases are quickly accessible in academic libraries. Researchers and students may seamlessly explore enormous knowledge libraries with rapid information retrieval.
- **Secure Data Transmission and Privacy Protection:** LiFi's limited signal range and resilience to interference increase security. It is a solid alternative for protecting academic database users and research data. Technology that uses visible light increases physical security, decreasing cyber dangers.
- **Real-time Collaboration and Research Advancements:** LiFi connection enables fast research collaboration. More efficient virtual meetings, data exchange, and joint projects foster academic creativity. LiFi helps collaborative research discoveries.
- **IoT Integration for Smart Libraries:** LiFi smoothly connects with IoT, creating intelligent environments in academic libraries. Bright lighting, sensor networks, and location-based services improve user experience. This integration makes libraries competent and responsive.
- **Energy Efficiency and Sustainability:** LiFi enhances energy efficiency in academic libraries using existing lighting infrastructure for data transmission. An eco-friendly Wi-Fi solution meets sustainability aims. As educational institutions promote green practices, LiFi promotes energy-efficient networking.

- **Advanced Learning Environments with AR/VR:** LiFi enables the development of advanced learning spaces in academic libraries. Interactive displays, augmented reality (AR), and virtual reality (VR) use high-speed connections to provide immersive teaching experiences. Students augment learning with unique content engagement.
- **Adoption of Future-Ready Technology:** LiFi drives the adoption of future-ready technology in academic libraries. Its interoperability with edge computing, AI, and blockchain puts educational institutions at the vanguard of technology.

## 5. LiFi's Impact on Academic Libraries' Location-Specific Information and Customized Services

LiFi guides academic libraries in their expansion, not merely for data transmission but also for creating intelligent, user-centric spaces by illuminating them with location-specific information, navigation, and personalized services. Through IoT sensor integration, LiFi supports exact location monitoring in libraries, providing precise, location-specific information for bookshelves, study rooms, and resource centres—a precision that underpins many location-based services. This technology enables academic libraries to create interactive maps and navigation systems, allowing users with LiFi-enabled devices to access real-time, interactive library maps that streamline navigation and improve the user experience, particularly for beginners and those seeking specific sections. As users navigate various library areas, LiFi delivers personalized services by notifying students of relevant events, seminars, and resources when they enter a particular topic zone, thereby customizing the library experience to individual interests and academic requirements. Working in conjunction with occupancy sensors, LiFi helps optimize smart study space allocation, enabling devices to discover empty study places in real-time, which minimizes user wait time and maximizes library resource utilization. The location-specific communication also facilitates quick access to relevant resources—for instance, a student standing in front of a bookshelf can receive brief information on related e-books or online resources directly on their device, creating a smooth transition between physical and digital knowledge repositories. Furthermore, LiFi enhances the attendance experience by notifying users of forthcoming events, workshops, or lectures as they travel around the library, boosting awareness and engagement through a tailored strategy. Finally, LiFi's location-specific features can integrate with augmented reality or virtual assistants, providing real-time information, instructions, and virtual guides depending on the user's location within the library, making the overall experience more engaging and futuristic.

## 6. Transforming Academic Libraries as Tech Hubs

Light Fidelity is changing academic libraries' learning settings as it revolutionizes connection. Using LiFi, these rooms become dynamic and technology-rich, enabling interactive displays, AR, and virtual learning environments.

- **Interactive Displays for Dynamic Engagement:** LiFi's fast connection enables the integration of interactive displays in academic libraries. Students and researchers may interact with digital information effortlessly in these dynamic knowledge centres. LiFi transforms libraries into learning places with interactive exhibitions of historical artifacts and collaborative digital whiteboards for brainstorming.
- **Augmented Reality (AR) for Immersive Experiences:** LiFi and AR technologies provide novel learning experiences. LiFi lets user devices seamlessly receive AR material, overlaying digital information on the actual world. Imagine students using AR glasses to explore a library section and get real-time book, historical, and multimedia information thanks to LiFi's high-speed and low-latency transmission.

- **Remote Collaboration with Virtual Learning Environments (VLEs):** LiFi integration bridges real and virtual areas. This programme allows students to access VLEs, attend virtual courses, and work on group projects from anywhere in the library, breaking down geographical boundaries and increasing library instruction.
- **Interactive Learning Modules and Educational Apps:** LiFi's fast data transmission enables smooth delivery of these tools. Devices provide personalized learning materials, interactive quizzes, and educational games to enhance learning. LiFi makes these materials accessible in academic libraries, facilitating varied learning methods.
- **Smart Bookshelves and Interactive Catalogs:** LiFi turns standard bookshelves into interactive centres. LiFi-enabled devices may provide real-time book information, reviews, and multimedia material on themes of interest. LiFi-powered interactive catalogues make resource discovery more fun and informative.
- **Collaborative AR Spaces for Group Projects:** LiFi enables collaborative AR spaces in the library. Virtual platforms allow students to manipulate 3D objects, share ideas, and study immersive. These joint efforts flow smoothly through a high-speed connection, encouraging collaboration and creativity.
- **Integrating LiFi with Educational Wearable:** LiFi integration with smart glasses or augmented reality headsets enhances interactive learning. Real-time information, instructional simulations, and virtual experiments are available. This integration blends the physical and digital worlds for an immersive education. Academic libraries may explore new possibilities with LiFi and other cutting-edge technology. Find out how LiFi and edge computing, AI, and blockchain may work together to become libraries and information centres.
- **Edge Computing for Rapid Data Processing:** Edge computing and LiFi's fast data transport are compatible. Academic libraries may improve real-time analytics by processing data at the network edge, enabling speedier insights into user behaviour, resource consumption, and operational efficiency. Edge computing makes LiFi apps more responsive, perfect for flexible library settings.
- **Artificial Intelligence for Personalized Learning:** LiFi and AI convert academic libraries into intelligent learning places. AI algorithms may customize learning experiences by analyzing user behaviour, preferences, and LiFi data. Imagine personalized resource suggestions, adaptive learning modules, and AI-powered virtual assistants improving library instruction.
- **Blockchain for Enhanced Security and Data Integrity:** Blockchain technology complements LiFi's limited signal range and provides decentralized security. Academic libraries may improve data security by merging LiFi and blockchain. This is crucial for academic record legitimacy, user data security, and library ecosystem digital asset tracking.
- **Smart Contracts for Resource Management:** Blockchain and LiFi work well together for resource management in academic libraries. Smart contracts automate and secure resource distribution, room booking, and access management. LiFi's location-specific features may activate smart contracts depending on user interactions, improving library efficiency.
- **Improved IoT Device connection:** LiFi's integration with edge computing and AI enhances academic library IoT device connection. LiFi's edge computing and high-speed data transmission can link smart sensors, interactive displays, and other IoT devices, creating a seamless ecosystem. This combination creates smart, responsive libraries.
- **Data Analytics for Informed Decision-Making:** Real-time data analytics using LiFi and edge computing empowers academic libraries with vital insights for informed decision-making. User habits, resource utilization, and engagement data may inform a strategy. Data-driven insights help libraries improve services, resource allocation, and user experience.
- **Integration of AR and VR Experiences:** LiFi and upcoming technologies provide immersive AR and VR experiences. Its high-speed LiFi networking lets AR and VR material stream smoothly, making library learning interactive. Interactive exhibitions, virtual simulations, and collaborative learning settings improve education.

## 7. Effects of Standardization on Academic Library LiFi Integration.

Standardization is crucial to the easy integration and broad acceptance of networking technologies like Light Fidelity, and these initiatives to standardize LiFi technology may significantly affect its incorporation into academic libraries. LiFi standardization promotes interoperability and device compatibility through uniform protocols and standards, which academic libraries need to seamlessly connect LiFi-enabled devices such as laptops and cell phones to the network, thereby reducing compatibility difficulties and uniting library devices. Additionally, standardization promotes uniform performance consistency across vendors, ensuring that academic libraries can deliver a dependable and consistent user experience—whether accessing digital materials or interactive displays—making LiFi in libraries reliable and constant. The development of security protocols and best practices through standardization overcomes data privacy issues and allows academic libraries to incorporate strong security measures, ensuring that user information, sensitive research data, and other important information are securely transferred across the library network. Standardization efforts are often conducted internationally through collaboration between industry stakeholders, regulatory bodies, and standardization organizations, which ensures that LiFi technology adheres to universal standards, facilitating global adoption within individual academic institutions and on a broader scale while promoting a consistent and harmonized approach to LiFi implementation worldwide. These initiatives also simplify integration with existing infrastructure by addressing interaction with components such as lighting and networking systems already present in academic libraries, offering recommendations and standards that streamline LiFi incorporation into libraries' technology ecosystems and increase uptake. Furthermore, standardization guarantees compliance with regulatory requirements, ensuring that LiFi deployments adhere to rules and creating a trustworthy and lawful environment for library data transit and storage. Finally, LiFi technology is future-proofed through standardization initiatives that create frameworks accepting breakthroughs and trends, allowing academic libraries to smoothly incorporate new features and capabilities as LiFi technology evolves, which is essential for libraries to remain ahead of technological advancements.

## 8. Promoting LiFi Technology in Academic Libraries

Academic libraries are actively promoting LiFi technology through a range of initiatives such as educational programmes, seminars, workshops, and awareness campaigns to highlight its potential in enhancing library services. Educational programmes including seminars, webinars, and expert talks help students, faculty, and staff understand LiFi’s features, applications, and advantages over traditional connectivity, while hands-on workshops allow users to experience LiFi-enabled devices and explore applications like high-speed data transfer, location-based services, and interactive learning environments. Libraries also organize LiFi showcase events with live demonstrations, interactive exhibits, and expert presentations to make the technology more visible and engaging. Collaborations with LiFi industry partners through guest lectures, sponsored events, and joint research further strengthen understanding of LiFi’s role in future connectivity. Awareness campaigns using posters, pamphlets, social media, and newsletters emphasize LiFi’s speed, security, and innovative uses, while training programmes equip library staff with the technical skills needed to support users and maintain LiFi services. Additionally, collaboration with academic departments integrates LiFi into teaching and research activities, encouraging its academic use. Despite these efforts, LiFi adoption in academic libraries faces challenges such as limited range and line-of-sight dependency, interference from ambient light sources, mobility constraints when users move between spaces, limited device compatibility and lack of standardization, high initial implementation costs, and potential security concerns that require strong encryption and cybersecurity measures.

## 9. Future LiFi Technology Trends and Developments

LiFi technology is anticipated to improve, removing constraints and opening new possibilities:

- **Increased Range and Improved Coverage:** Research and development may enhance LiFi signal range and coverage, making it more useful in library situations.
- **Improved Mobility Support:** LiFi may be enhanced to support mobile devices, guaranteeing continuous connection for users in library areas.
- **5G Network Integration:** LiFi and 5G networks may synergize to improve connection and solve device mobility and coverage issues.
- **Standardization for Interoperability:** Standardization initiatives will likely ensure interoperability and compatibility across LiFi-enabled devices from different vendors.
- **Smart Infrastructure and IoT Integration:** LiFi integration with IoT devices may enhance resource management, real-time tracking, and personalized services in smart libraries.
- **Hybrid Networks and Seamless Handovers:** The future may have hybrid networks that smoothly switch between LiFi and other wireless technologies, maintaining ongoing consumer connections.
- **Research on LiFi Applications in Specialized Areas:** Research organizations and libraries might investigate LiFi applications in specialized areas, such as labs, where accurate and secure communication is crucial.
- **Advancements in Visible Light Communication (VLC):** Visible Light Communication (VLC) advancements may enhance LiFi signal quality, data transfer speeds, and performance under adverse lighting circumstances. Academic libraries must remain up-to-date on LiFi technological advances, undertake rigorous feasibility evaluations, and prepare for future integration.

## 10. Conclusion

The demand for mobile data traffic in tube environments in the future will be too great for current radio frequency (RF) communication solutions to handle. As such, different electromagnetic spectrum bands like visible light and millimetre waves are being investigated by researchers (Wang et al., 2023). Integrating Light Fidelity (LiFi) technology inside academic libraries is a promising and dazzling way to improve connection, communication, and user experience. As we light the way to the future of information transmission, LiFi shines as a beacon of innovation that may alter academic learning settings. Libraries with LiFi provide quicker and more dependable access to digital resources. Students, instructors, and researchers may collaborate more efficiently, and research procedures will be streamlined. Improved connection keeps academic libraries at the forefront of technological progress, meeting the changing demands of scholars.

The security characteristics of LiFi, based on visible light communication, make information sharing more safe and private. Keeping sensitive data and intellectual property secure is essential in academia. LiFi's tolerance to electromagnetic interference makes it a cutting-edge technology that can fulfil academic institutions' strict security needs. The dazzling voyage with Light Fidelity in academic libraries guarantees fast, secure, and efficient connection. As we use light to transport information, we enter a new age of technology integration that improves learning, stimulates cooperation, and preserves academic institutions as knowledge and innovation centres. LiFi illuminates a future where information is as seamless as light transmission.

## References

Haas, Harald. (2017). LiFi is a Paradigm-Shifting 5G Technology. *Reviews in Physics*. *3*. 26–31. 10.1016/j.revip.2017.10.001.

Haas, H. Yin, L, Wang Y. & Chen, C. (2016). "What is LiFi?" *Journal of Lightwave Technology*, 34(6), 1533-1544.

Shaaban, R., & Faruque, S. (2020). “Cyber security vulnerabilities for outdoor vehicular visible light communication in secure platoon network: Review, power distribution, and signal to noise ratio analysis”. *Physical Communication*, 40. doi:10.1016/j.phycom.2020.101094

Verma, T., Shrivastava, S., Dwivedi, U. D., & Kothari, D. P. (2023). “Transfer learning for resource allotment in dynamic hybrid Wi-Fi/ LiFi communication systems”. *Optics Communications*, 546.

Wang, X., Gan, C., Wu, S., Chen, Y., & Chen, Y. (2023). “Rate-stabilized and load-balanced handover management for subway hybrid LiFi and Wi-Fi networks”. *Optics Communications*, 549.